\documentclass{article}
\usepackage[T1]{fontenc}
\usepackage[utf8]{inputenc}
\usepackage{ismir} % Remove the "submission" option for camera-ready version
\usepackage{amsmath,cite,url}
\usepackage{graphicx}
\usepackage{color}

\usepackage{tabularx}
\usepackage{multirow}
\usepackage{amsfonts}
\usepackage{xspace}
\usepackage{lipsum}
\usepackage{tcolorbox}
\usepackage{adjustbox}
\usepackage{booktabs}
\usepackage{makecell}
\usepackage{enumitem}
\usepackage{listings}
\usepackage{pgfplots}
\pgfplotsset{compat=1.18}
\newcommand{\dataset}{$\texttt{Not that Groove}$\xspace}

\title{Zero-Shot Symbolic Music Editing as a Reasoning Task for Large Language Models}

\oneauthor
  {Li Zhang}
  {Drexel University\\\texttt{Harry.Zhang@drexel.edu}}

\def\authorname{Li Zhang}

\begin{document}

\maketitle

\begin{abstract}
While recent advancements in AI music generation have predominantly focused on direct audio synthesis, these systems suffer from inherent rigidity, limiting their utility for professional music producers who require granular, highly malleable creative control. Symbolic music (e.g., MIDI) resolves this constraint by providing editable note-level parameters, yet the natural progression to instruction-driven symbolic music \textit{editing} remains critically under-explored due to a severe scarcity of paired instruction-MIDI datasets. In this paper, we bypass this data bottleneck by formalizing zero-shot symbolic music editing as a structured reasoning task. We introduce a novel text-based ``drumroll'' notation that translates musical mechanics into a spatial, syntax-driven grid, empowering off-the-shelf Large Language Models (LLMs) to logically deduce and apply complex edits to drum grooves using only zero-shot prompting. To rigorously evaluate this paradigm, we propose \dataset, a comprehensive benchmark comprising thousands of drum grooves paired with specific, descriptive, and stylistic natural language instructions. Crucially, to overcome the prohibitive cost and subjectivity of human musical evaluation, we introduce a scalable, domain-informed automated unit-testing framework that symbolically verifies whether an edited groove satisfies the core constraints of the user's request. Our extensive experiments across eight state-of-the-art LLMs demonstrate the high efficacy of this approach, with the top-performing model achieving a 68\% success rate on our automated unit tests. Furthermore, listening tests confirm that our programmatic unit tests align highly with the subjective judgments of professional musicians, establishing a robust, data-efficient, and scalable foundation for the future of controllable AI music production.\footnote{The reader is encouraged to listen to the demos in the accompanying materials matching the speaker icons \raisebox{-0.3\height}{\includegraphics[width=0.5cm]{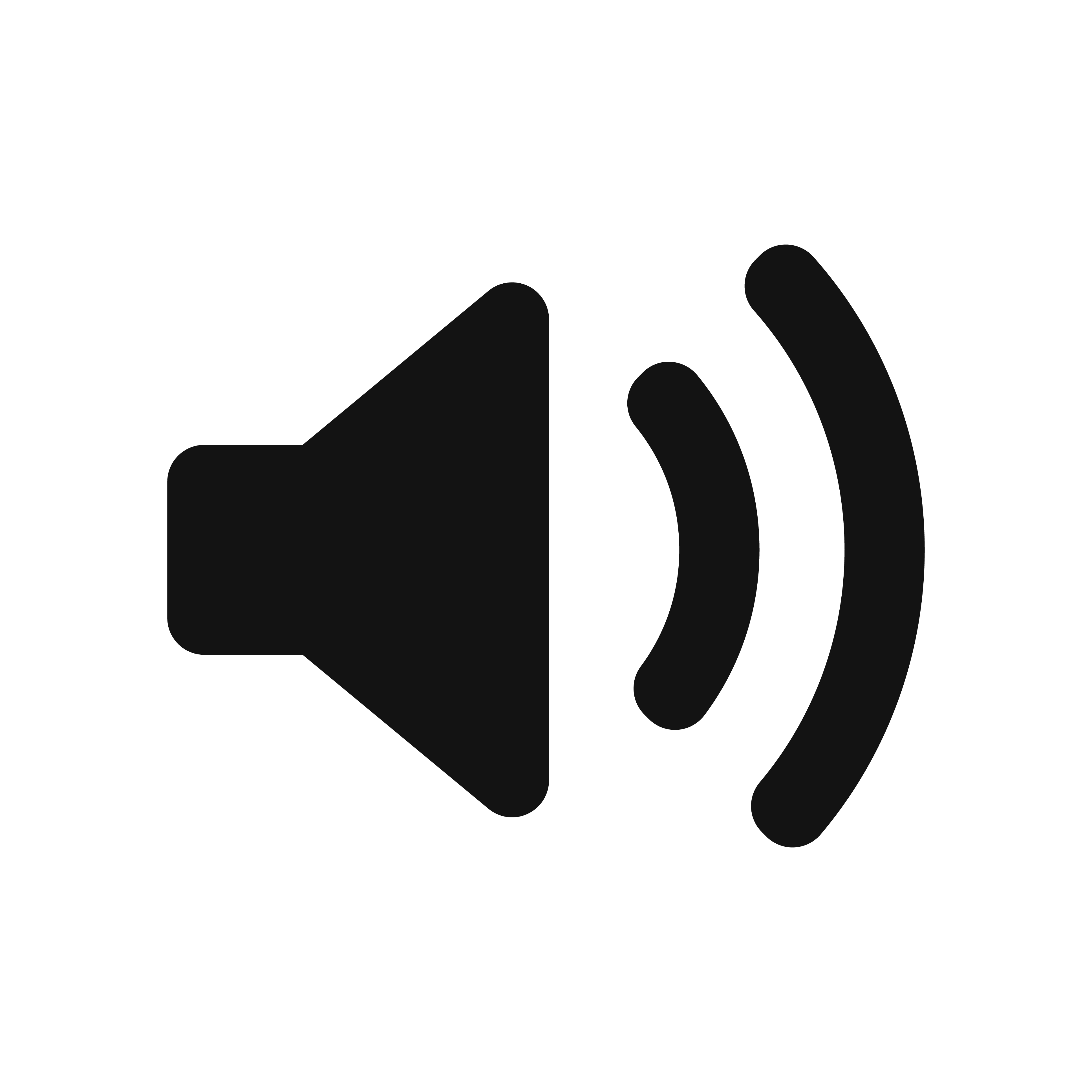}} throughout the paper.}
\end{abstract}

\section{Introduction}

Music generation has seen much development along transformers and large language models (LLMs).
Most existing systems tackled \textbf{music audio generation} \cite{dhariwal2020jukebox,agostinelli2023musiclmgeneratingmusictext,copet2024simplecontrollablemusicgeneration} given textual queries.
While made publicly popular by commercial products such as Suno\footnote{\url{https://suno.com/}}, these tools have seen little use for music producers beyond a novelty \cite{tencer2024ai}.
Apart from concerns of quality, one major reason is the lack of creative control, as generated audio can hardly be edited, fine-tuned, or adapted (analogous to presenting a made meal to a chef).
Despite ongoing efforts of \textbf{music audio editing} based on textual instructions \cite{zhang2024musicmaguszeroshottexttomusicediting,tsai2024audiopromptadapterunleashing,hou2025editingmusicmelodytext,lan2024highfidelitytextguidedmusic}, a line of research that may be much more appealing to music producers is \textbf{symbolic music generation}  \cite{hadjeres2017deepbachsteerablemodelbach,openai2019musenet,zeng-etal-2021-musicbert,zhang2023language,bhandari2024text2midigeneratingsymbolicmusic,huang2024symbolicmusicgenerationnondifferentiable}.
These system output sheet music or Musical Instrument Digital Interface (MIDI), which can be readily used in performance or production (analogous to presenting ingredients to a chef).
Even so, if a producer is dissatisfied with the generated MIDI, it would be ideal for them to be able to verbally describe desired changes to be automatically implemented (e.g., make this sound more jazzy).
Unfortunately, efforts on \textbf{symbolic music editing} are missing in literature, likely due to the lack of data that pairs instructions and symbolic music edits, contrary to more ample data for music audio.

\begin{figure}[!t]
    \centering
    \includegraphics[width=\linewidth]{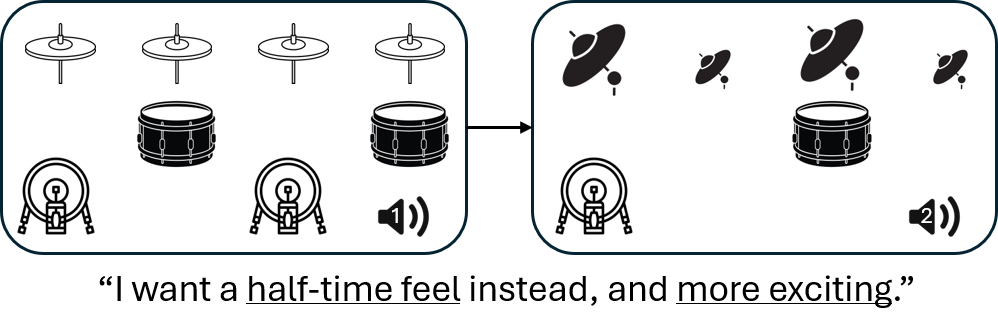}
    \caption{A pictorial illustration of gpt-4.1's edit on the \dataset dataset.
The original groove (left) plays the hi-hat on all 4 beats, the snare drum on beat 2 and 4, and the kick drum on beat 1 and 3. The edited groove (right) changes the constant hi-hat hits to dynamic crash cymbal hits, typically increasing excitement of the music;
it halves the frequency of the snare drum and the kick drum, resulting in a half-time feel.}
    \label{fig:thumbnail}
\end{figure}

\begin{figure*}[!t]
    \centering
    \includegraphics[width=\textwidth]{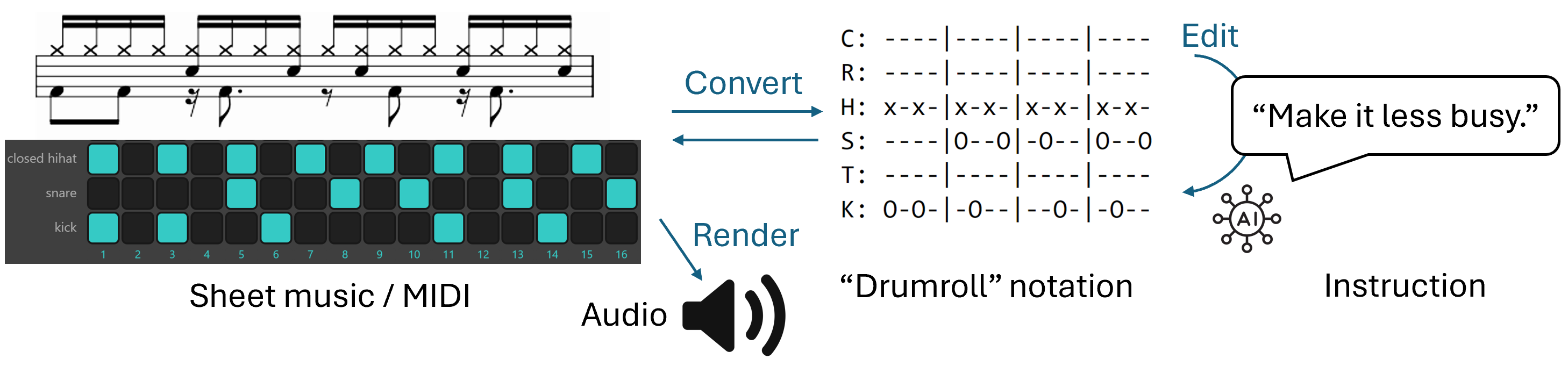}
    \caption{A given 1-bar drum groove in represented as a MIDI file or some sheet music in converted into a drumroll notation.
Given a user instruction, tn LLM generates an edited drum groove also in the drumroll notation, which can be converted back to MIDI or sheet music in a rule-based manner, which can be rendered as audio using samples or by performing on an actual drum set.}
    \label{fig:pipeline}
\end{figure*}

To address this challenge, we show it is possible to edit symbolic music based on textual instructions via creative zero-shot prompting of LLMs to cast the task into a reasoning problem (Figure~\ref{fig:thumbnail}).
We prove this concept through a case-study of editing drum grooves which underpin popular music \cite{witek2014syncopation}.
The recipe for success is finding an appropriate representation that works well with LLM inference, so that LLMs may treat the music editing task as a reasoning task over structured data.
For evaluation, we introduce the \dataset dataset with thousands of drum grooves paired with textual instructions annotated by a professional drummer and music producer.
The instructions are diverse in nature, including specific ones, stylistic ones, descriptive ones, etc., and clearly labeled.
As music is inherently subjective, providing ground-truth labels of edits is likely futile.
Hence, a key innovation of the dataset is having a unit test associated with each example, which can symbolically evaluate if an edited drum groove meets minimal requirements.
Out of 8 LLMs we experiment with, the best model gpt-4.1 makes 68\% passing edits.
We next prove that the unit tests highly corresponds with musician's judgment with a listening test that shows that 72\% are in fact correct and musical edits that a drummer would propose.
Overall, we provide system that takes a drum groove and an edit instruction as input, and outputs an edited drum groove in terms of sheet music, MIDI, and audio.

In summary, our main contributions are:
\begin{itemize}
    \item We present the first work to show that zero-shot symbolic music editing is possible by making an analogy to a reasoning task that LLMs excel at.
    \item We introduce the \dataset dataset, providing a novel benchmark for a case study featuring annotated instructions and automated unit tests for evaluating symbolic drum edits.
\end{itemize}

\section{Representation}
\label{sec:representation}

We study a particular kind of symbolic music, the composition of a drum set, that is core to modern popular music.
To interface LLMs for understanding and generation, we use the transposed \textit{drumroll} representation inspired by the success of previous work \cite{zhang2023language}.
An example can be seen in the rightmost of Figure~\ref{fig:pipeline}. Each line corresponds to an instrument on a drum set.
We consider 6 instruments: kick drum, snare drum, toms, hi-hat, crash cymbal, and ride cymbal.
Each instrument can be performed differently, resulting in different articulations.
A line represents a bar of 4/4 time signature containing 4 beats, each containing 4 16th note.
Each character thus represents a 16th note, grouped by 4 into beats separated by $|$.
A character is $-$ if the instrument is not played on that note.
When played, the character denotes the articulation which varies by instruments.
Altogether, one such representation constitutes a \textit{one-bar groove}, which is the basis of many genres of music.
A representation like can be converted to a MIDI format, which can be rendered as audio.
While MIDI and audio are more expressive, we prefer the drumroll notation based on its minimal nature and analogy to natural language, as each note is akin to an alphabet, each beat is akin to a word, and each bar is akin to a sentence.
We assert that this representation works best with LLMs especially without specific training.

\section{Experimental Setup}
\subsection{Formulation}
We consider the task of editing symbolic music based on a natural language request.
Concretely, a model is given $g_o$, an original one-bar groove in a drumroll notation (e.g., the example in Figure~\ref{fig:pipeline}), and an instruction $i$ describing a user request (e.g., ``I want it to sound heavier'').
The model should output a new groove $g_e := \text{LLM}(g_o,i)$.
The desiderata of a good $g_e$ is one that not only implements the change requested(e.g., by changing the hi-hat hits to cymbal hits and halve the frequency of the kick and snare drum) in a musical way.

\subsection{Dataset}
Evaluating an edited art form such music, visual art, or creative text often require human judgment that introduces great cost and subjectivity.
To overcome this, we propose the \dataset dataset. We start with a manually labeled subset as the development set that contains 31 tuples of $g_o$ and $i$ annotated by a professional drummer.
The instructions $i$ span many labeled categories, such as specific (e.g., ``I'd like a Cymbal hit and a Kick on the first note.''), descriptive (e.g., ``This beat is too basic.''), or stylistic (e.g., ``This should have a more jazzy vibe.'').
As music is subjective in nature, annotating any ground-truth $g_e$ would be misguided.
While most existing work defers to costly human listening tests, we propose a \textbf{unit test} $t$ and its arguments $\{arg_j\}$ for each example.
A unit test can be a conjunction or disjunction of multiple unit tests.
The unit test automatically checks if the edited one-bar groove symbolically fulfill certain minimal constraints.
For example, a \textit{specific} instruction of ``I don't want any kick drum in the first beat'' should never result in any groove with kick drums played in any of the 4 16th notes in the first beat, regardless of many possibilities of the edited groove itself.
Similarly, a \textit{descriptive} instruction of ``The hi-hats are too busy'' should never result in any groove with even more hi-hat hits, however they are arranged.
Or, a \textit{stylistic} instruction of ``Make it sound more funky'' should almost always result in a groove with some notes on the back-beats (those that are not the first note of each beat).
Therefore, the unit tests are annotated to be a necessary but insufficient condition of a good edit.
Nevertheless, we complement this automatic evaluation with human ratings of other professional musicians on best-performing models.

\begin{table}
    \centering
    \begin{tabular}{llll} \toprule
         & Dev & Test & Example \\ \midrule
        Total & 31 & 1,116 & \\ 
        specific & 22 & 1,023 & No cymbal on beat 4 \\
        descriptive & 6 & 83 & A bit less busy on hi-hats \\
        stylistic & 3 & 11 & Sound like thrash metal \\ \bottomrule
    \end{tabular}
    \caption{Number of examples by types of instructions in the \dataset dataset.}
    \label{tab:statistics}
\end{table}

To scale up the data size, we rewrite some instructions as templates (e.g., ``I'd like a [inst1] hit and a [inst2] on the first note.'').
Instantiated with all possible instruments, the instructions are paired with 8 seed original grooves covering 8 genres to form a test set of 1,116 tuples of $g_o$ and $i$.
While we evaluate models on the full dataset, we by default report and discuss results on the manually labeled subset due to considerations of efficiency and data quality.
The statistics of the dataset is shown in Table~\ref{tab:statistics}.\footnote{The relative lack of descriptive and stylistic instructions is due to the difficulty of defining unit tests for a particular feel or style.
We are currently expanding these instructions.} 

An example in the \dataset dataset has the following component.
First, the original groove $g_o$ in the drumroll notation may be:
\begin{lstlisting}
Basic 2 and 4 pop groove.
K: O---|----|O---|----
S: ----|O---|----|O---
H: x---|x---|x---|x---
T: ----|----|----|----
C: ----|----|----|----
R: ----|----|----|----
\end{lstlisting}
Next, a templated instruction may be:
\begin{lstlisting}
I'd like a @inst0@ hit and @ins1t@ in the very beginning.
\end{lstlisting}
This instruction is accompanied by a unit test $t$ and templated arguments ${arg_j}$:
\begin{lstlisting}
t := have_inst_on_note("@i0@", 0) && have_inst_on_note("@i1@", 0)
\end{lstlisting}
where the unit test is defined as:
\begin{lstlisting}
def have_inst_on_note(inst, pos):
    beats = drum_dict.get(inst, [])
    notes = [note for beat in beats for note in beat]
    # If the 16th note at the position is not '-', return True
    if notes[pos] != '-':
        return True
    return False
\end{lstlisting}
The templated instruction is instantiated once in the development set, while instantiated by all combinations of 
two instruments in the test set:
\begin{lstlisting}
I'd like a crash cymbal hit and kick in the very beginning.
\end{lstlisting}
the unit test call is also instantiated:
\begin{lstlisting}
t := have_inst_on_note("C", 0) && have_inst_on_note("K", 0)
\end{lstlisting}
which checks the edited groove in the drumroll notation if the first character after `C: '  and that after `K: ' is a valid articulation such as `O' or `o'.

\subsection{Modeling and Prompting}
We use a zero-shot prompt to familiarize LLMs with the drumroll notation. For example:
\begin{lstlisting}
You will compose some drum beats for a song. First, let's learn about a drum notation.
A bar of drum beats may look like this:
@@@
K: O---|----|O---|----
S: ----|X--o|----|O---
H: x---|x---|x---|x---
T: ----|----|-O--|---o
C: O---|----|----|----
R: O---|----|----|----
@@@
Each line corresponds to an instrument on a drum set:
K: Kick drum
S: Snare drum
H: Hihat
T: Toms
C: Crash cymbal
R: Ride cymbal
Each character in a line represents a 16th note.
Each four characters separated by | constitute a beat. Note that there are 16 characters, not counting the |, because there are 16 16th notes in a bar which constitute 4 beats.
Each character is - if the instrument is not played on that note.
When played, the character denotes the articulation which varies by instruments.
K: O is a hard hit, while o is a soft hit
S: O is a hard hit, while o is a soft open hit on the head;
additionally, X and x are hard and soft sidestick hits
H: O is a hard open hit, while o is a soft open hit;
additionally, X and x are hard and soft closed hits
T: O is a hard hit, while o is a soft hit
C: O is a hard hit, while o is a soft hit
R: O is a hard open hit on the bell, while o is a soft open hit on the bell;
additionally, X and x are hard and soft closed hits on the bow
\end{lstlisting}

Next, we provide them with the original groove represented as a drumroll and the instruction for the edit.
\begin{lstlisting}
You are given the following drum groove.
@@@
K: O---|----|O---|----
S: ----|O---|----|O---
H: x---|x---|x---|x---
T: ----|----|----|----
C: O---|----|----|----
R: ----|----|----|----
@@@
\end{lstlisting}

Next, we provide the edit instruction and prompt the model to start generation.
For example:
\begin{lstlisting}
You received the following edit request.
"I don't want any kick."
You will now edit this drum groove and generate a new one in the above notation.
You are free to show your thought process, but only the final groove should be between @@@ which will be used.
\end{lstlisting}

With this input, the LLM generates an edited groove as a drumroll. A reasonable edited drum groove may be:
\begin{lstlisting}
@@@
K: ----|----|----|----
S: ----|O---|----|O---
H: x---|x---|x---|x---
T: ----|----|----|----
C: ----|----|----|----
R: ----|----|----|----
@@@
\end{lstlisting}
Note that not only the kick drums but also the crash cymbal hit is removed because conventionally, crash cymbals on the first beat go with kick drums.

If the generated drumroll representation is malformed (e.g., does not have exactly 6 instruments, or missing some notes), the evaluation is immediately aborted.
Otherwise, it is input into the corresponding unit test to perform automatic evaluation.
Next, the drumroll notation is converted into MIDI with a default setting of 120 BPM and 4/4 time signature.
Finally, the MIDI including different articulations and velocity is rendered into an audio clip by triggering and mixing samples from a drum sound library.
The audio clips are played during the listening tests. The procedure of evaluation is shown in Figure~\ref{fig:pipeline}.

We experiment with an array of open- and closed-source state-of-the-art LLMs of various sizes, including GPT4.1-mini, GPT-4.1-nano\footnote{\url{https://openai.com/index/gpt-4-1/}} using OpenAI's API, DeepSeek-R1-Distill-Llama-70B, DeepSeek-R1-Distill-Llama-80B \cite{deepseekai2025deepseekr1incentivizingreasoningcapability}, and QwQ-32B\footnote{\url{https://qwenlm.github.io/blog/qwq-32b/}} using H100 GPUs.

\begin{figure}[t!]
\small
    \centering
\begin{tikzpicture}
    \begin{axis}[
        width=\columnwidth, height=5cm,
        ybar=0pt,
        bar width=12pt,
        ymin=0, ymax=100,
        xtick=data,
        xticklabel style={rotate=20, anchor=east, font=\scriptsize},
        symbolic x coords={gpt-4.1-nano, gpt-4.1-mini, DeepSeek-R1-8B, DeepSeek-R1-70B, QwQ-32B},
        legend style={at={(0.7,0.98)}, anchor=north, legend columns=2, font=\scriptsize},
        ylabel={\% passed unit test},
        nodes near coords,
        every node near coord/.append style={font=\tiny},
        enlarge x limits={abs=1cm}
    ]

    \addplot coordinates {(gpt-4.1-nano, 22.3) (gpt-4.1-mini, 67.7) (DeepSeek-R1-8B, 25.8) (DeepSeek-R1-70B, 35.4) (QwQ-32B, 64.5)};
    \addplot coordinates {(gpt-4.1-nano, 42.9) (gpt-4.1-mini, 79.1) (DeepSeek-R1-8B, 7.5) (DeepSeek-R1-70B, 41.4) (QwQ-32B, 61.3)};
    \legend{development set, test set}
    \end{axis}
\end{tikzpicture}
    \caption{The percentage of generated drum grooves that pass the unit tests.}
    \label{fig:unit_test_results}
\end{figure}
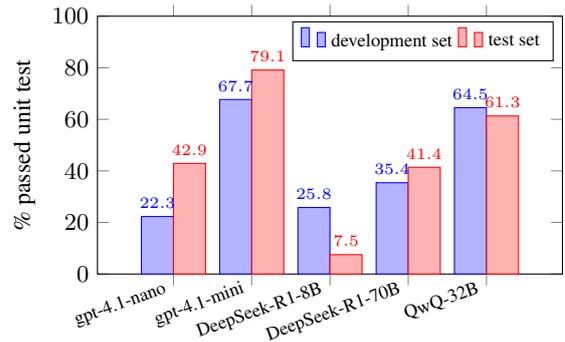

\section{Results}

\subsection{Unit Test}
The performance of all LLMs on both the development and set set, measured by whether the edited drum grooves pass the unit test, is shown in Figure~\ref{fig:unit_test_results}.
For both the gpt-4.1 family and the DeepSeek-R1 family of models, the performance positively correlates with the model size.
Interestingly, QwQ-32B outperforms the 70B DeepSeek-R1 models, only second to gpt-4.1-mini.
As QwQ-32B is shown to excel at coding an reasoning, it is not a far cry to compare the task of symbolic music editing to a task of structured reasoning.
Even so, music is art beyond just symbols and should not only be evaluated using algorithms.

\subsection{Listening Test}
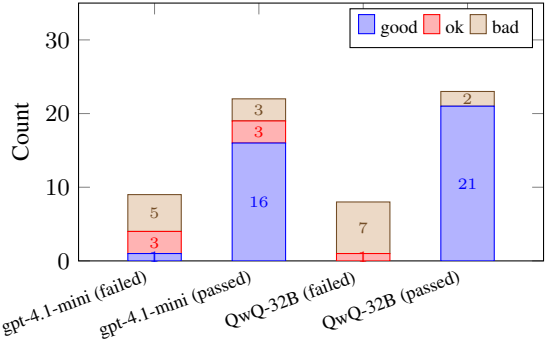
\begin{figure}[t!]
    \centering
    \small
    \begin{tikzpicture}
    \begin{axis}[
        width=\columnwidth, height=5cm,
        ybar stacked,
        bar width=20pt,
        ymin=0, ymax=35,
        xtick=data,
        xticklabel style={yshift=-5pt, rotate=20, anchor=east, font=\scriptsize},
        symbolic x coords={gpt-4.1-mini (failed), gpt-4.1-mini (passed), QwQ-32B (failed), QwQ-32B (passed)},
        legend style={at={(0.78,0.98)}, anchor=north, 
legend columns=-1, font=\scriptsize},
        ylabel={Count},
        nodes near coords,
        every node near coord/.append style={font=\tiny},
        enlarge x limits={abs=1cm}
    ]

    \addplot coordinates {(gpt-4.1-mini (failed), 1) (gpt-4.1-mini (passed), 16) (QwQ-32B (failed), 0) (QwQ-32B (passed), 21)};
    \addplot coordinates {(gpt-4.1-mini (failed), 3) (gpt-4.1-mini (passed), 3) (QwQ-32B (failed), 1) (QwQ-32B (passed), 0)};
    \addplot coordinates {(gpt-4.1-mini (failed), 5) (gpt-4.1-mini (passed), 3) (QwQ-32B (failed), 7) (QwQ-32B (passed), 2)};
    \legend{good, ok, bad}
    \end{axis}
\end{tikzpicture}
    \caption{Human musician's judgment on the development set of drum grooves generated by best models, grouped by if they passed the automatic unit tests.}
    \label{fig:human_results}
\end{figure}

To evaluate the models more realistically, we ask a professional drummer and producer to listen to all edited grooves in the development set and annotate one of the following labels: \textit{bad}, if the edit would by no means pass as done by a human musician, \textit{ok}, if the edit could be done by some musician but might not be the best choice, and \textit{good} if the edit is faithful to the instruction while staying musical.
Figure~\ref{fig:human_results} shows that the unit test provided in \dataset has a high true positive rate (counting both \textit{good} and \textit{bad} as positive) of 89\% and a high true negative rate of 94\%.
It can therefore be considered a reliable automatic metric to evaluate edited drum grooves whenever human listening test is unavailable.
While gpt-4.1-mini has a slightly higher pass rate of unit tests than QwQ-32B, the human musician prefers the edits of QwQ-32B slightly more frequently among the 31 examples in the development set (21 \textit{good} over 16, 7 \textit{bad} over 8).
Both models demonstrate strong ability to edit drum grooves based on instructions.

\subsection{Qualitative Study}
\label{sec:qualitative}

\begin{figure}[t!]
    \centering
    \includegraphics[width=\columnwidth]{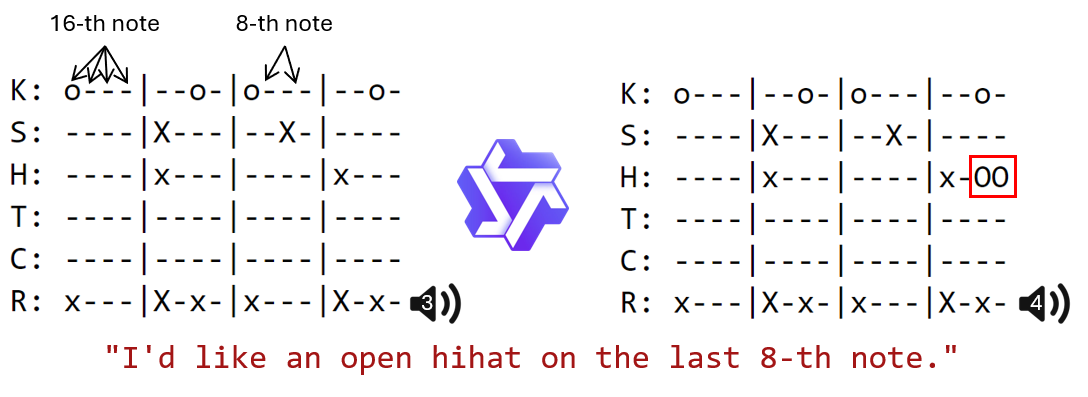}
    \includegraphics[width=1.27\columnwidth]{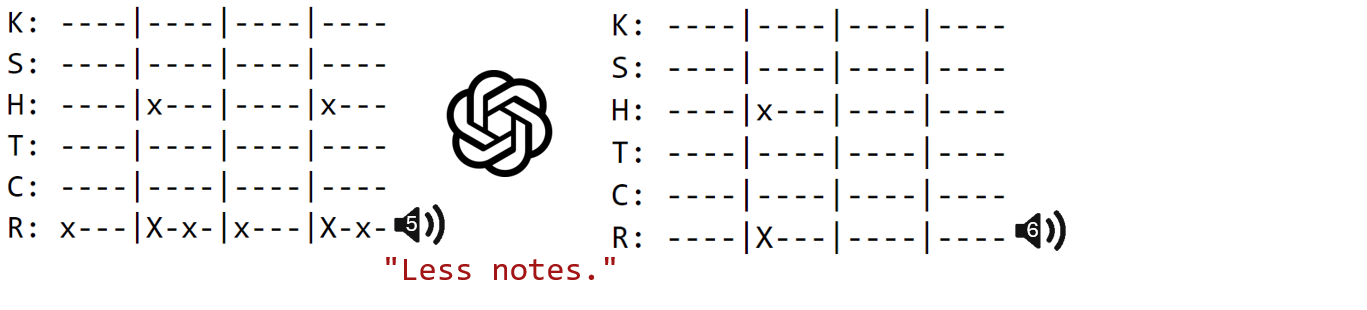}
    \caption{Two examples of a successful and unsuccessful edit.}
    \label{fig:example1}
\end{figure}

We closely look at two edited grooves, one successful and one unsuccessful (Figure~\ref{fig:example1}).
We examine how the LLMs reach them, why they are good or bad, an how they can be applied in a realistic music production scenario.

\subsubsection{Example 1: A Successful Bossa Nova Edit} 
The model is provided with a bossa nova style groove, with a specific request to add an open hi-hat in the last 8th note, a common technique in the arrangement of drums.
Recall from Section~\ref{sec:representation} that each character (delimited by `-') represents an 16th note.
Per basic music theory, an 8th note is the span of every two 16th notes.
Therefore, the last 8th note of the hi-hat refers to the last two characters on the third row.
While simple for human even with minimal music training, QwQ-32B is the only LLM that generates the correct edit. By inspecting its reasoning trace, we can observe exactly how the model tackles symbolic music editing by casting it as a multi-step reasoning problem over a structured grid.

The model first correctly grounds itself in the rules of the drumroll notation provided in the prompt, linking the musical instruments to their textual representations:
\begin{lstlisting}
The user wants an open hihat on the last 8th note. First, I need to recall the drum notation details.
Each line represents an instrument, and each character is a 16th note.
The hihat (H) uses O for open hits and X/x for closed.
\end{lstlisting}

Next, the model identifies the core ambiguity of the user's natural language request (``last 8th note'') and attempts to map this temporal musical concept into a spatial coordinate within the representation. It debates whether the user means the last 8th note of a single beat or the entire measure:
\begin{lstlisting}
The user mentioned the "last 8th note," which is a bit ambiguous.
Since each beat is four 16th notes, an 8th note is two of those.
Looking at the current H line: H: ----|x---|----|x---. Each | separates beats.
The bar has four beats, so the last beat is the fourth one.
The last 8th note of the entire bar would be the 14th and 15th 16th notes?
Wait, maybe they mean the last 8th note of each beat? Or the very last 8th note in the bar?
\end{lstlisting}

After exploring these interpretations, the model systematically breaks down the 16-character sequence beat-by-beat to locate the exact string indices required for the edit:
\begin{lstlisting}
Beat 1: ---- (all four 16th notes silent except maybe the first? Wait, no: the first part is "----" so all four 16th notes of beat 1 have no hihat.
...
Beat4: x, -, -, -
So the hihat is only on the first 16th note of beat2 and beat4.
...
So the last 8th note is the last two 16th notes of the bar.
Since the current H line's last beat is x---, which is x followed by three dashes.
So the last two 16th notes (positions 15 and 16) are both dashes.
\end{lstlisting}

With the target coordinates identified (positions 15 and 16 on the `H` line), the model formulates the exact string manipulation. Crucially, it treats the edit as a strict constraint satisfaction problem, explicitly checking that substituting the dashes with `O` respects the 16-character mathematical constraint of the bar:
\begin{lstlisting}
But the user wants an open hit on the last 8th note, which is two 16th notes.
So replacing the third and fourth with O. So the last beat becomes x-OO.
Wait, but let's check the total length.
Each line must have 16 characters (four beats of four each).
Let's count:
Original H line: 
Beat1: ---- (4) ... Beat4: x--- (4). Total 16.
After changing beat4 to x-OO, that's still four characters.
So the new H line would be:
H: ----|x---|----|x-OO
\end{lstlisting}

Finally, the model verifies that no unintended changes were made to the other instruments before outputting the final groove. This step-by-step breakdown illustrates that LLMs do not need to inherently ``hear'' or ``feel'' music to edit it effectively; by converting the musical structure into a semi-structured text syntax, the artistic request transforms into a solvable sequence of reading comprehension, spatial reasoning, string manipulation, and constraint checking.

\subsubsection{Example 2: A Failed Swing Jazz Edit} 
The model is provided with a swing jazz groove, with a simple descriptive request to decrease the number of notes.
The edit proposed by gpt-4.1-mini does exactly that, passing the unit test.
However, the model removes all but two notes: a hi-hat hit and a ride cymbal hit only on the second downbeat. This way, the groove does not sound like a musical jazz groove any more. By examining the reasoning trace for this type of failure, we can observe the pitfalls of treating music generation purely as a symbol-reduction task without acoustic or stylistic grounding.

Unlike the rigorous spatial reasoning that led to success in the previous example, the model here falls into the trap of over-optimizing for the explicit text constraint. It begins by correctly parsing the descriptive instruction, but immediately adopts an overly literal and aggressive mathematical strategy:
\begin{lstlisting}
The user wants to "decrease the number of notes."
Let me analyze the current note density.
The Ride cymbal (R) has a standard swing pattern: 10 notes.
The Hi-hat (H) is played on beats 2 and 4: 2 notes.
The Snare (S) has some syncopated comping: 4 notes.
Total notes = 16. 
To satisfy the request to "decrease," I need to output a groove with significantly fewer than 16 notes.
\end{lstlisting}

Instead of preserving the structural pillars of a swing groove, such as the defining ``spang-a-lang'' rhythm on the ride cymbal or the steady hi-hat ``chick'' on the off-beats, the model treats all characters in the string as equally expendable. It proceeds to mechanically prune the grid without musical hierarchy:
\begin{lstlisting}
Let's clear the S line completely to reduce clutter. S: ----|----|----|----
The R line is very busy. I will remove the swing subdivisions and just keep one hit to establish the ride cymbal sound.
Let's keep the H line only on the second downbeat, and remove the fourth beat to further decrease the note count.
Now, H is just on beat 2 (position 5), and R is also on beat 2 (position 5).
New Total Notes = 2. 
2 is strictly less than 16. The decrease constraint is successfully met.
\end{lstlisting}

Finally, the model confidently validates its own output against the unit test, completely unaware that the resulting sparse matrix has entirely lost its genre identity:
\begin{lstlisting}
Let me double check the constraints. 
Are there fewer notes? Yes (2 < 16).
Are there exactly 16 characters per line? Yes.
Are the instrument labels correct? Yes.
Outputting the new groove.
\end{lstlisting}

This failure mode highlights a critical limitation of zero-shot LLMs in creative domains. While the model efficiently satisfies the discrete, logical bounds of the unit test (i.e., $N_{edited} < N_{original}$), it lacks the intrinsic musical ``common sense'' to balance mathematical reduction with stylistic preservation. Without specific, programmatic definitions of what makes a groove a \textit{swing jazz} groove, the model blindly treats musical notes as mere data points to be minimized.

These two examples show that LLMs possess existent but imperfect reasoning ability to apply specific edits using basic music theory even without any training.

\section{Conclusion and Limitations}

We presented a novel paradigm for AI-driven music generation by formalizing zero-shot symbolic music editing as a structured reasoning task. By introducing a text-based drumroll notation, we successfully interfaced the spatial reasoning and constraint-satisfaction capabilities of Large Language Models with the structural rules of music theory. To overcome the subjective evaluation of musical edits, we contributed the \dataset dataset and a scalable unit-testing framework that highly correlates with professional musicians' judgments. However, our case study is currently restricted to discrete, 1-bar drum grooves. The proposed drumroll representation cannot yet capture uncommon instruments, nuanced articulations (e.g., rimshots, bell hits), complex sub-divisions (e.g., triplets), human feel (e.g., swung or laid-back feel), or long-term structural dependencies.

Building upon this foundation, future work must address the representational bottlenecks required to scale this methodology to continuous polyphonic instruments, full multi-instrument ensembles, or mixing projects, which will likely necessitate modular architectures. Furthermore, to mitigate the stylistic failure modes observed in purely zero-shot LLM edits, future systems could integrate automated planning and neurosymbolic methods. By pairing the deep natural language understanding of LLMs with rigorous, rule-based planning algorithms, we can enforce strict musical logic and stylistic guardrails, ultimately empowering producers with precise and highly malleable creative control over entire multi-track compositions.

% -------------------------------------------------------------------------
% NOTE ON APPENDICES:
% The ISMIR 2026 guidelines explicitly forbid appendices in the submitted PDF.
% The appendix code below has been retained from your ACL manuscript so you 
% don't lose the text, but you MUST remove or comment it out before compiling 
% your final PDF submission. You can supply this content in supplementary materials.
% -------------------------------------------------------------------------

%\section{Licensing}
%The \dataset dataset is licensed under CC BY-SA 4.0.
%All models used in this work are done so in a way that comply with their licenses.

%\section{Annotator Information}
%The annotator that contributed to the \dataset dataset is the author themselves, who is a professional drummer and content creator who runs a video channel of over 60,000 subscribers and 10,000,000 views.
%The annotator was provided with strict guidelines, informed that the data collected will be for research purposes, and compensated fairly.

\bibliography{anthology, custom}

\end{document}